\documentclass[floatfix,aps,twocolumn,nofootinbib,superscriptaddress,preprintnumbers,pra,10pt]{revtex4-1}

\usepackage[utf8]{inputenc}
\usepackage{float}
\usepackage{colortbl}
\usepackage{amsmath,amssymb,amsfonts, bm,bbm,slashed}
\usepackage{dsfont} 
\usepackage{hyperref}
\usepackage{graphicx}
\usepackage{enumitem}
\usepackage{arydshln}
\usepackage{mathtools}
\usepackage{bbold}
\usepackage{braket}
\usepackage{multirow}
\usepackage{soul}
\usepackage[dvipsnames]{xcolor}

\usepackage{feynmp-auto}
\makeatletter
\g@addto@macro\bfseries{\boldmath}
\makeatother

\newcommand{\be} {\begin{equation}}
\newcommand{\ee} {\end{equation}}
\newcommand{\bea} {\begin{eqnarray}}
\newcommand{\eea} {\end{eqnarray}}
\newcommand{\no} {\nonumber}
\newcommand{\ba} {\begin{array}}
\newcommand{\ea} {\end{array}}
\newcommand{\gsim}{\lower.7ex\hbox{$\;\stackrel{\textstyle>}{\sim}\;$}}
\newcommand{\lsim}{\lower.7ex\hbox{$\;\stackrel{\textstyle<}{\sim}\;$}}
\renewcommand{\Re}{{\rm Re}}

\newcommand{\cL}{\mathcal{L}}
\newcommand{\cC}{\mathcal{C}}
\newcommand{\cO}{\mathcal{O}}

\newcommand{\mlq}{m_U^2}

\definecolor{violet}{cmyk}{0,1,0,0.2}

\allowdisplaybreaks

\begin{document}

\preprint{ZU-TH-42/21}

\title{LFU violations in leptonic $\tau$ decays and $B$-physics anomalies}

\author{Lukas Allwicher}
\email{lukall@physik.uzh.ch}
\affiliation{Physik-Institut, Universit\"at Z\"urich, CH-8057 Z\"urich, Switzerland}
\author{Gino Isidori}
\email{isidori@physik.uzh.ch}
\affiliation{Physik-Institut, Universit\"at Z\"urich, CH-8057 Z\"urich, Switzerland}
\author{Nud{\v z}eim Selimovi{\'c}}
\email{nudzeim@physik.uzh.ch}
\affiliation{Physik-Institut, Universit\"at Z\"urich, CH-8057 Z\"urich, Switzerland}

\begin{abstract}
\vspace{5mm}
We present a complete analysis of Lepton Flavor Universality (LFU) violations in leptonic $\tau$ decays 
in motivated models addressing the $B$-physics anomalies, based on the  
$SU(4)\times SU(3)\times SU(2)\times U(1)$ gauge group. We show that the inclusion of 
vector-like fermions, required by $B$-physics data, leads to sizable modifications of the leading-log 
results derived within an Effective Field Theory approach. In the motivated parameter-space region 
relevant to the $B$-physics anomalies, the models predict a few per-mil decrease of the effective 
$W$-boson coupling to $\tau$, within the reach of future experiments.
\vspace{3mm}
\end{abstract}

\maketitle

\section{Introduction}
 The per-mil level tests of Lepton Flavor Universality in $\tau$ decays~\cite{Pich:2013lsa}
are among the most stringent constraints on physics beyond  the Standard Model (SM)
close to the electroweak scale.  
These tests are particularly interesting and challenging in view of the hints of LFU violations 
reported in semileptonic $B$ decays, the so-called $B$-physics anomalies, 
whose evidence has been rising over the years~\cite{LHCb:2021trn,deSimone:2020kwi}.
Already in the early attempts to address the $B$  anomalies, these constraints provided 
serious limitations on the proposed new physics (NP) explanations  (see e.g.~Ref.~\cite{Greljo:2015mma}).
In this context, a key observation was made in Ref.~\cite{Feruglio:2016gvd,Feruglio:2017rjo}:
even if $\tau$ decays are not affected at the tree level by NP models 
 addressing the $B$ anomalies, the latter necessarily affect $\tau$ decays at the one-loop level. 
More precisely,  NP models addressing $b \to c \tau \bar \nu$ anomalies 
 via a modification of the left-handed (semileptonic) $b$-decay amplitudes, 
lead to sizable one-loop corrections in $\tau$ decays. 
The leading-log contribution is model independent, and is determined by 
the RG evolution of the semileptonic operators in SM Effective Field Theory (SMEFT)~\cite{Jenkins:2013wua}.
The size of the discrepancy between data and theory in 
$b \to c \tau \bar \nu$ transitions naturally implies LFU violations in 
purely leptonic $\tau$ decays at the few per-mil level.

So far, all analyses of these effects have been based on leading-log Effective Field Theory (EFT) results.
However,  finite one-loop corrections arising from matching conditions at the NP scale
might be relevant, both given the 
large values of the effective couplings in the most motivated NP models and
the small separation between  electroweak and NP scales.
 This is particularly true in ultraviolet (UV) complete models which predict a non-trivial spectrum for the heavy states.

In this paper we analyse such finite corrections in the so-called 4321 models, 
i.e.~models based on the gauge group 
$SU(4)\times SU(3)\times SU(2)\times U(1)$~\cite{DiLuzio:2017vat,Bordone:2017bld,Greljo:2018tuh,DiLuzio:2018zxy,Fuentes-Martin:2020bnh,Fuentes-Martin:2020hvc,Fuentes-Martin:2020pww},  
where the color group, $SU(3)_c$, is the diagonal (unbroken) subgroup of $SU(4)\times SU(3)$. 
The spontaneous symmetry breaking $4321\to{\rm SM}$ leads to a 
massive vector leptoquark (LQ), $U_1$, which is a very effective 
tree-level mediator for the $B$ anomalies~\cite{Barbieri:2015yvd,Buttazzo:2017ixm}.
We focus in particular on flavor non-universal  
4321 models~\cite{Bordone:2017bld,Greljo:2018tuh,Fuentes-Martin:2020bnh,Fuentes-Martin:2020pww,Fuentes-Martin:2020hvc}, where only third-generation fermions are charged under $SU(4)$, 
providing a natural justification for the flavor structure of the $U_1$ couplings~\cite{Barbieri:2015yvd}.

The one-loop structure of 4321 models, which naturally include also vector-like fermions and scalar fields, 
 has been investigated in~\cite{Fuentes-Martin:2019ign,Fuentes-Martin:2020luw,Fuentes-Martin:2020hvc}.
Recent phenomenological analyses~\cite{Cornella:2021sby} suggest a non-trivial hierarchy in the 
spectrum of the different NP states, with heavy vectors and relatively light vector-like fermions.
As we shall see, the latter can play a relevant role in the LFU breaking effects in $\tau$ decays.

\section{EFT expressions for the LFU ratios}

The observables we are interested in are the purely leptonic LFU ratios
\bea
   \left|g^{(\tau)}_e/g^{(\mu)}_e\right|^2 &\equiv & \frac{\Gamma(\tau\to e\nu\bar{\nu})}{\Gamma(\mu\to e\nu\bar{\nu})}\left[\frac{\Gamma_{\rm SM}(\tau\to e\nu\bar{\nu})}{\Gamma_{\rm SM}(\mu\to e\nu\bar{\nu})}\right]^{-1},\quad
\label{eq:one}
\eea
with $|g^{(\tau)}_\mu/g^{(\mu)}_e|^2$ and $|g^{(\tau)}_\mu/g^{(\tau)}_e|^2$ defined in complete analogy.
By construction, these ratios are expected to be equal to one within the SM. Their  current experimental 
world averages can be found in Ref.~\cite{Pich:2013lsa}.

We work under the assumption that the new degrees of freedom modifying $\tau$ (and $\mu$) decays 
occur above the electroweak scale. Under this assumption, we can describe 
the relevant NP contributions via the so-called low-energy EFT (LEFT) Lagrangian,
obtained by integrating out  new degrees of freedom and heavy SM fields ($W$,  $Z$, $t$, and $H$):
\begin{align}
\cL_{\rm LEFT}=-\frac{2}{v^2}\sum_k\,\cC_k \cO_k\,.
\end{align}
Using the notation of Ref.~\cite{Jenkins:2017jig}, 
where the RG structure of $\cL^{\rm LEFT}$ can also be found,
the operators contributing at the tree level to pure leptonic decays are 
\begin{align}
    [\cO_{\nu e}^{V,LL}]_{\alpha\beta\gamma\delta} &= (\bar{\nu}_{L}^{\alpha}\gamma_\mu\nu_{L}^{\beta})(\bar{e}_{L}^{\gamma}\gamma^\mu e_{R}^{\delta})\label{eq:LVLL}\,,\\
    [\cO_{\nu e}^{V,LR}]_{\alpha\beta\gamma\delta} &= (\bar{\nu}_{L}^{\alpha}\gamma_\mu\nu_{L}^{\beta})(\bar{e}_{R}^{\gamma}\gamma^\mu e_{R}^{\delta})\,.
\end{align}
Since the SM decay amplitude is purely left-handed (LH) and we work under the hypothesis of 
small NP corrections, quadratic NP effects and 
the contributions of the right-handed (RH) operators can be safely neglected. 
To a very good accuracy, we can write 
\bea
    R_{\beta\alpha} &\equiv&  \frac{\Gamma(\ell_\beta\to\ell_\alpha\nu\bar{\nu})}{\Gamma_{\rm SM}(\ell_\beta\to\ell_\alpha\nu\bar{\nu})} \equiv 1 + \delta R_{\beta\alpha} 
    \label{eq:dR}  \\ 
&   \approx&  1 +  2\, \Re[\cC_{\nu e}^{V,LL}]^{\rm NP}_{\alpha\beta\beta\alpha}\,,
\eea
where we have used $[\cC_{\nu e}^{V,LL}]^{\rm SM}_{\alpha\beta\beta\alpha} = 1$,
up to tiny scale-independent electroweak corrections that we can safely neglect.

The evaluation of the leptonic LFU ratios thus reduces to the evaluation of the NP contributions to 
$\cC_{\nu e}^{V,LL}$, at the electroweak scale. To achieve this goal we need to match the explicit NP 
model onto the SMEFT Lagrangian at the heavy scale, which we normalise as 
\begin{align}
\cL_{\rm SMEFT}=-\frac{2}{v^2}\sum_k\, C_k O_k\,,
\end{align}
run down to the electroweak scale, and finally match the SMEFT onto the LEFT. 
Starting from the leading SMEFT semileptonic operators relevant to the 
$B$-physics anomalies, namely
\bea
[O_{\ell q}^{(1)}]_{\alpha \beta i j } &=& (\bar \ell_{L}^{\alpha}   \gamma_{\mu}  \ell_{L}^{\beta})
(\bar q_{L}^{i}    \gamma^{\mu}  q_{L}^{j} )\, , \no\\ {}
[O_{\ell q}^{(3)}]_{\alpha \beta i j } &=& (\bar \ell_{L}^{\alpha}  \sigma^I \gamma_{\mu}  \ell_{L}^{\beta})
(\bar q_{L}^{i}  \sigma^I  \gamma^{\mu}  q_{L}^{j} )\, , 
\eea
performing a tree-level matching, and considering 
the leading-log contribution in the RG evolution of the SMEFT operators, 
leads to~\cite{Feruglio:2016gvd,Feruglio:2017rjo}
\be  
[\cC_{\nu e}^{V,LL}]^{\rm NP-LL}_{\alpha\beta\beta\alpha}   =  
-  \frac{ m_t^2 N_{\rm c}}{ 4\pi^2 v^2} \log\frac{\mu^2} {m_t^2} \sum_{\gamma=\alpha,\beta}
[C_{\ell q}^{(3)}]_{\gamma\gamma 33}~.
\label{eq:C_leading}
\ee
where $N_{\rm c} =3$ is the number of colors and $\mu$ denotes the UV matching scale.

In this paper we go one step forward in precision, both using 
one-loop SMEFT-LEFT matching conditions at the low scale, 
and taking into account the high-scale one-loop
matching of the 4321 model onto the SMEFT. This way we systematically 
control not only the leading-log corrections but also all the relevant finite terms
(at the same order in the perturbative expansion in terms of the LQ coupling $g_U$).
Proceeding this way, Eq.~(\ref{eq:C_leading}) gets modified as follows
\bea
&&    [\cC_{\nu e}^{V,LL}]^{\rm NP-full}_{\alpha\beta\beta\alpha}  =  - 2 \sum_{\gamma=\alpha,\beta}
[C_{H\ell}^{(3)}]_{\gamma\gamma}(\mu)  + \no\\
   &&\quad    + [C_{\ell\ell}]_{\alpha\beta\beta\alpha} + [C_{\ell\ell}]_{\beta\alpha\alpha\beta}+\label{eq:SMEFT-LEFT}  \no \\
    &&\quad  -  \frac{ m_t^2 N_{\rm c}}{ 8\pi^2 v^2} \sum_{\gamma=\alpha,\beta}
[C_{\ell q}^{(3)}]_{\gamma\gamma 33} \left(1+2\log\frac{\mu^2}{m_t^2}\right)~. 
\eea
Here $C_{H\ell}^{(3)}$, $C_{\ell\ell}$ are the coefficients  of the operators 
\bea
    [O_{H\ell}^{(3)}]_{\alpha\beta} &=& (\bar{\ell}^\alpha\gamma_\mu \sigma^I \ell^\beta) (H^\dagger i \overleftrightarrow{D^\mu}\sigma^I H)\,,  \\ {}
    [O_{\ell\ell}]_{\alpha \beta \gamma\delta } &=& (\bar \ell_{L}^{\alpha}   \gamma_{\mu}  \ell_{L}^{\beta})
    (\bar \ell_{L}^{\gamma}   \gamma^{\mu}  \ell_{L}^{\delta})~,
\eea
obtained by the one-loop matching of the NP model onto the SMEFT.
In section~\ref{sect:matching} we derive the  explicit expressions of these coefficients 
in terms of masses and couplings of the heavy fields in the 4321 model.

\section{The model}
\subsection{Simplified version: SM fermions only}
It is convenient to consider first a simplified version of the model with minimal 
fermion content. In this limit only three chiral fermions are charged under $SU(4)$:
they can be identified with the third generation of SM fermions supplemented by a RH neutrino 
($\nu_R^3$).
The transformation properties of these chiral fields under the complete 4321 gauge group
is~\cite{Fuentes-Martin:2020hvc}
\begin{alignat}{2}
    \psi_L &= (q_L^3\ \ell^3_L)^T &&\sim (\textbf{4, 1, 2})_{0}\,,\\
    \psi^+_R &= (t_R\ \nu_R^3)^T &&\sim (\textbf{4, 1, 1})_{1/2}\,,\\
    \psi^-_R &= (b_R\ \tau_R)^T &&\sim (\textbf{4, 1, 1})_{-1/2}\,,
\end{alignat}
where $t_R$, $b_R$, and $\tau_{R}$ have been identified with the corresponding mass-eigenstates,
while $q_L^3$ and $\ell_L^3$
denote the quark and lepton doublets. For the sake of concreteness, we assume 
$q_L^3$ and $\ell_L^3$ are aligned to the down-quark and 
charged-lepton mass basis, respectively (hence $\ell_L^3 \equiv \ell_L^\tau$).
We comment on the impact of this assumption at the end of Section~\ref{sect:Cll}.
These quantum-number assignments give rise to the following interaction 
between SM fermions and the vector LQ:
\bea
\Delta \cL_U  &= & 
\frac{g_U}{\sqrt{2}} U_1^\mu J^U_\mu + \mathrm{h.c.}~,  \qquad  \no \\
 J^U_\mu &=&  \bar q_{L}^{\,3} \gamma_{\mu}  \ell^3_{L} 
+\bar b_{R} \gamma_{\mu} \tau_{R} 
+ \bar t_{R} \gamma_{\mu} \nu^3_{R}\,.
\label{eq:JU0}
\eea
The tree-level exchange of the  $U_1$ field leads to 
\be
[C_{\ell q}^{(3)}]_{\tau\tau 33} = \frac{1}{2} C_U\,, \qquad 
C_U  = \frac{g_U^2 v^2}{4 \mlq}~.
\label{eq:LQtree}
\ee
In this simplified version of the model, the SM fermions of the first and second generation,
which are singlets under $SU(4)$, do not couple to the $U_1$.

\subsection{Inclusion of vector-like fermions}
\label{sect:2B}
In order to generate a non-vanishing coupling of the $U_1$ to second generation fermions, 
the field content is enlarged including an additional $SU(4)$-charged 
left-handed fermion
\begin{alignat}{2}
    \chi_L & = (Q_L^\prime \ L_L^\prime)^T &&\sim (\textbf{4, 1, 2})_0\,,
\end{alignat}
and a corresponding RH partner ($\chi_R$) with the same SM
quantum numbers.\footnote{For the purpose of this analysis, we do not need to distinguish the 
case where $\chi_R$ transform as a $(\textbf{4, 1, 2})_0$, from the case where $\chi_R$
 indicates two separate fields ($Q_R$ and $L_R$) transforming as $(\textbf{1, 3, 2})_0$ and $(\textbf{1, 1, 2})_0$, 
 respectively~\cite{Fuentes-Martin:2020hvc}.}

After the $4321\to{\rm SM}$ symmetry breaking, the effective mass terms 
in the Lagrangian lead to two vector-like (VL) states ($Q$ and $L$, with different masses), 
whose LH components mix with  the LH chiral fermions.
The inclusion of the new $SU(4)$-charged fields modifies 
the LH current in Eq.~(\ref{eq:JU0})  into
\be
 \bar q_{L}^{\,3} 
 \gamma_{\mu} \ell^3_{L} 
\ \to\  \left(\bar q_{L}^{\,3}\  \bar Q^\prime_{L} \right) W  
 \gamma_{\mu} \left(\ba{c} \ell^3_{L}  \\  L^\prime_{L}  \ea \right)\,,
 \label{eq:Wmatrix}
\ee
where $W$ is a $2\times 2$ unitary matrix with a potentially large mixing angle 
controlling the mixing of  the exotic fermions and third-generation  chiral fermions.
The states $Q_L^\prime$ and $L_L^\prime$ are not  mass eigenstates due to the 
additional (small) mixing with second-generation chiral fermions. Expressing them 
in terms of the the mass-eigenstates leads to 
\bea
    Q_L'  &=& c_Q Q_L - s_Q q_L^2\,, \no\\
    L_L'  &=& c_L L_L - s_L \ell_L^2\,, 
    \label{eq:Qp}
\eea
with $s_{L,Q} \ll 1$ and $c_{L,Q} =\sqrt{1- s^2_{L,Q}}\approx 1$. The states orthogonal 
to those in Eq.~(\ref{eq:Qp}) are the would be second-generation chiral fermions in absence of 
mixing that, by construction, do not interact with the $U_1$ field 
 (see Ref.~\cite{Fuentes-Martin:2020hvc} for more details).

In principle, the model could be modified adding also heavy fermions which could mix with the 
$SU(2)_L$-singlet chiral fermions $\psi^{\pm}_R$.
This addition,   which implies a modification of  the RH current in Eq.~(\ref{eq:JU0}), 
  has no direct impact on the amplitudes we are interested in. However, 
  it might have an indirect impact changing the best-fit value of $C_U$ resulting 
from the global fit of the $B$ anomalies~\cite{Cornella:2021sby}. After the inclusion of both sets of heavy fermions,
the LQ current in Eq.~(\ref{eq:JU0}), expressed in terms of mass-eigenstates, assumes the generic form
\be
J^U_\mu = \sum_{ \lbrace \psi^{i}_L \rbrace }  \beta_L^{\psi^i \psi^j}  \bar \psi^i_{L} \gamma_{\mu}  \psi^j_{L} +
\sum_{ \lbrace \psi^{i}_R \rbrace }  \beta_R^{\psi^i \psi^j}  \bar \psi^i_{R} \gamma_{\mu}  \psi^j_{R}~. 
\label{eq:JUgen}
\ee

In addition to modifying the  LQ current, the field $\chi_L$ couples to the right-handed SM fermions and the 
SM Higgs field via a  (4321 invariant) Yukawa interaction
\be
\Delta\cL_Y =  Y^\prime_- \bar{\chi}_L \psi^-_R H +  Y^\prime_+ \bar{\chi}_L \psi^+_R \tilde{H} + {\rm h.c. }  \,,
\ee
where $ \tilde{H}=i\sigma_2H^\dagger$.
Expressing the latter in term of mass-eigenstates, leads to the following interactions between $u_R$, $d_R$, and the 
heavy fermions
\be
\Delta\cL_Y  \supset 
c_Q Y_- \bar{Q}_L  d_R^3 H +  c_QY_+ \bar{Q}_L  u^3_R \tilde{H}  + \rm h.c. \,,
\ee
where the difference between $Y^\prime_\pm$ and $Y_\pm$ takes into account the possible mixing in the RH sector. 
Note that $\Delta\cL_Y$ induces also a contribution to the effective SM Yukawa interaction:
\be
\Delta\cL_Y  \supset 
-  s_Q Y_- \bar{q}^2_L  d_R^3 H -  s_Q  Y_+ \bar{q}^2_L  u^3_R \tilde{H}  + \rm h.c. \,.
\ee
This implies $|Y_+ | \sim  y_t |V_{cb}|/ s_Q  \gg |Y_-|$, where $y_t$ is the top-quark Yukawa coupling 
and $V_{ij}$ denote the matrix elements of the Cabibbo-Kobayashi-Maskawa matrix.

\section{One-loop matching conditions}
\label{sect:matching}

\subsection{$U_1$ + SM fields}

\begin{figure}[t]
\centering
\includegraphics[width=0.35\textwidth]{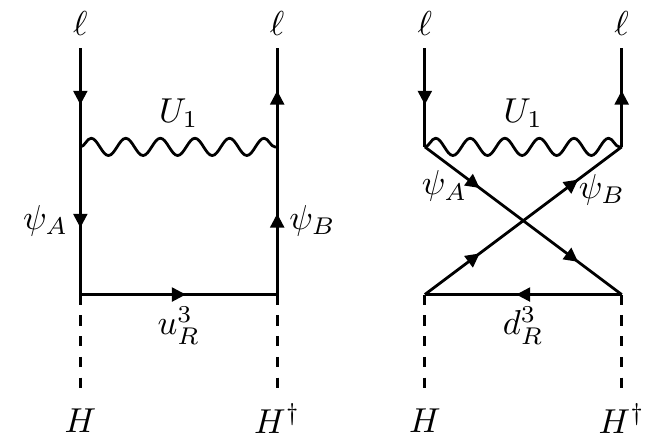}
\caption{One-loop diagrams in the full theory 
contributing to the correlation function
$\braket{\ell \bar{\ell} H H^\dagger}$.
 $\psi_{A,B}$ generically denote left-handed fermions charged under $SU(4)$.}
\label{fig:VL_SM_Box}
\end{figure}

We first derive the matching condition to $C_{H\ell}^{(3)}$ in the simplified model with only SM fermions.
To this purpose,  we consider the off-shell Green's function 
\be
 \braket{\ell_\beta^b(0) \bar{\ell}_\alpha^a(0) H^c (q) H^{\dagger d}(-q)}, 
\label{eq:GG}
\ee
where $a,b,c,d$ are $SU(2)_L$ indices, and all momenta are taken incoming. The one-loop diagrams 
in the UV  theory contributing to this correlation function are 
shown in  Fig.~\ref{fig:VL_SM_Box}. In this case $\psi_{A,B}$  is identified with $q^3_L$
and, since we neglect the bottom Yukawa coupling, only the diagram on the left contributes.

Since we are interested only in the $SU(2)_L$-triplet component of the  correlation function,
we concentrate on the part of the amplitude proportional to the factor 
$(\sigma^I)^a_b (\sigma^I)_c^d$ (which is omitted in the amplitudes reported below). 
Computing the amplitude in the full theory in the limit $m^2_t \ll |q^2| \ll m^2_U$
leads to
\bea
 	\left[\mathcal{A}_{\rm UV,0}\right]_{\tau\tau} &=&
 	-\frac{4iN_{\rm c}}{16\pi^2 v^2}\ |y_t|^2 [C_{\ell q}^{(3)}]_{\tau\tau 33} \times \no\\
 	&& \times    \log\left(\frac{\mlq}{-q^2}\right)\ \bar{v}(0) \slashed{q}P_L u(0)\,, 
\eea
where $P_L = (1-\gamma_5)/2$, 
with $C_{\ell q}^{(3)}$ given in Eq.~(\ref{eq:LQtree}). As can be seen, the amplitude exhibits an infrared 
singularity, which is regularized by $q^2\not=0$.

\begin{figure}[t]
\centering
\includegraphics[width=0.35\textwidth]{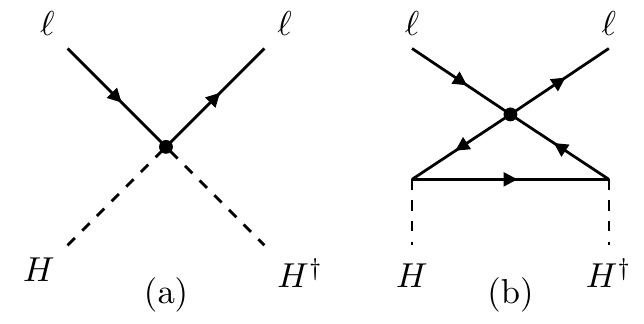}
\caption{Tree-level and one-loop diagrams in the SMEFT contributing 
to the correlation function $\braket{\ell \bar{\ell} H H^\dagger}$.}
\label{fig:EFTloop}
\end{figure}

In order to perform the matching we need to compute the same correlation function in the SMEFT.
At the one-loop level, the relevant diagrams are shown in Fig.~\ref{fig:EFTloop}.
 The amplitude corresponding to the diagram in Fig.~\ref{fig:EFTloop}a is simply
\begin{equation}
	\left[\mathcal{A}_{\rm EFT,a}\right]_{\tau\tau}=-\frac{4i}{v^2} [C_{H\ell}^{(3)}]_{\tau\tau}  \bar{v}(0) \slashed{q}P_L u(0)\,,
\end{equation}
while the amplitude generated by the diagram in Fig.~\ref{fig:EFTloop}b, in the limit 
$m^2_t \ll |q^2|$, reads
\bea
	\left[\mathcal{A}_{\rm EFT,b}\right]_{\tau\tau}&&= -\frac{4iN_{\rm c}}{16\pi^2 v^2}  |y_t|^2 [C_{\ell q}^{(3)}]_{\tau\tau 33} 
	\times \no\\
	&& \times   \left(1+\log\frac{\mu^2}{-q^2}\right)\  \bar{v}(0) \slashed{q}P_L u(0)\,. 
\eea
As expected, $\mathcal{A}_{\rm EFT,b}$ exhibits  the same infrared structure of $\mathcal{A}_{\rm UV}$.
By imposing the  relation
\be
 \mathcal{A}_{\rm EFT,a}(\mu) = \mathcal{A}_{\rm UV} - \mathcal{A}_{\rm EFT,b}(\mu) 
 \label{eq:mathing}
\ee
with  $\mathcal{A}_{\rm UV} = \mathcal{A}_{\rm UV,0}$ 
we determine the matching condition 
\be
	[C_{H\ell}^{(3)}]_{\tau\tau} = - \frac{1}{16\pi^2} N_{\rm c} |y_t|^2 [C_{\ell q}^{(3)}]_{\tau\tau 33}  \left(1+\log \frac{\mu^2}{\mlq}\right)\,.
\ee
As a consistency check, from this result we deduce 
that the running of $C_{H\ell}^{(3)}$ due to $C_{\ell q}$ is 
\begin{equation}
	16\pi^2 \mu \frac{\partial }{\partial \mu}  [C_{H\ell}^{(3)}]_{\tau\tau}  = - 2 N_{\rm c} |y_t|^2 [C_{\ell q}^{(3)}]_{\tau\tau 33} \,,
\end{equation}
which matches the known result in \cite{Jenkins:2013wua}.

\subsection{UV amplitude in the complete model}

We now proceed evaluating the contributions to the UV amplitude 
from the additional heavy states present in the complete model.
In this case the contribution of the SM fermions proceeds as above,
but the tree-level expression for  $C_{\ell q}^{(3)}$ changes because 
of the modified LQ  current in  Eq.~(\ref{eq:Wmatrix}). In particular,
one gets  
\be
[C_{\ell q}^{(3)}]_{\tau\tau 33} = \frac{1}{2} |W_{11}|^2 C_U\,, \quad 
[C_{\ell q}^{(3)}]_{\mu\mu 33} = \frac{1}{2} |W_{12}|^2 s_L^2 C_U\,. 
\label{eq:LQtreeF}
\ee

The VL fermions lead to two additional terms. 
The diagrams in  Fig.~\ref{fig:VL_SM_Box} where 
both $\psi_{A}$ and $\psi_{B}$ are identified with VL fermions, 
and those where only one of them is a VL fermion, the other 
being $q_L^3$. In the first case both diagrams are non vanishing and yield\footnote{Here, we give the amplitudes for third generation leptons in the external states. The analogous result for the second generation leptons can be obtained by replacing $W_{i1} \to -s_L W_{i2}$.}
\bea
 &&   \left[\mathcal{A}_{\rm VL,1}\right]_{\tau\tau}  = \frac{2 i N_{\rm c}}{16\pi^2}  \frac{g_U^2 |W_{21}|^2 c_Q^4}{4\mlq} \times  \no\\
  && \qquad \times    (|Y_+|^2 + |Y_-|^2) B_1(x_Q)\  \bar{v}(0) \slashed{q}P_L u(0)\,,
\eea
where $x_Q = m_Q^2/\mlq$, with $m_Q$ being the VL quark mass, and 
\begin{equation}
    B_1(x_Q) = \frac{1-x_Q+\log x_Q}{(1-x_Q)^2}\,.
\end{equation}
In the second case, neglecting all the SM Yukawa couplings except for $y_t$, 
the result is 
\bea
      \left[\mathcal{A}_{\rm VL,2}\right]_{\tau\tau}  &&  =\frac{2 i N_{\rm c}}{16\pi^2}  \frac{g_U^2 c_Q^2}{4\mlq} 2{\rm Re}(W_{11}^* W_{21} Y_+^* y_t)  \times \no\\
   && \times 
 B_0(x_Q)\ \bar{v}(0) \slashed{q}P_L u(0)\,,
\eea
with $B_0(x_Q) = \log x_Q / (1-x_Q)$. 

The above results hold in the Feynman gauge. In this gauge we need to take into account 
also the contributions from diagrams of the type in Fig.~\ref{fig:VL_SM_Box}, 
with the $U_1$ replaced by the corresponding Goldstone boson (GB).
The GB amplitudes with one or two SM fermions are vanishing, while the 
one with two VL fermions yields
\bea
  &&   \left[\mathcal{A}_{\rm GB}\right]_{\tau\tau}  =- \frac{2 i N_{\rm c}}{16\pi^2}  \frac{g_U^2 |W_{21}|^2 c_Q^4}{4\mlq}  \times  \no\\
 && \qquad  \times  (|Y_+|^2 + |Y_-|^2) B_2(x_Q)\ \bar{v}(0) \slashed{q}P_L u(0)\,,
\eea
where
\begin{equation}
	B_2(x_Q) = \frac{x_Q-x_Q^2+x_Q^2\log x_Q}{4(1-x_Q)^2}\,.
\end{equation}

Finally, the contributions where the $U_1$ is replaced by the corresponding radial excitation (Higgs mode,
with mass $m_{h_U}$)  should also taken into account. In this case we find
\bea
  &&     \left[\mathcal{A}_{\rm R}\right]_{\tau\tau}  = -\frac{2 i N_{\rm c}}{16\pi^2}  \frac{g_U^2 |W_{21}|^2 c_Q^4 \tan^2\beta}{4\mlq}
  \times  \no\\
 && \qquad  \times   (|Y_+|^2 + |Y_-|^2) B_2(x_Q^R)\ \bar{v}(0) \slashed{q}P_L u(0)\,,
\eea
with $x_Q^R = m_Q^2/m_{h_U}^2$  and $\tan\beta=\omega_1/\omega_3$, where $\omega_1$ and $\omega_3$ are the vacuum expectation values of the scalar fields mediating the 4321 $\to$ SM breaking~\cite{Fuentes-Martin:2020hvc}. We neglect model-dependent 
contributions involving quartic scalar couplings of the radial modes. 

\subsection{Complete matching for $C_{H\ell}^{(3)}$}
We are now in the position to sum all the contributions and obtain the matching conditions
for both $[C_{H\ell}^{(3)}]_{\tau\tau}$ and $[C_{H\ell}^{(3)}]_{\mu\mu}$ in the complete model.
Proceeding as in  Eq.~(\ref{eq:mathing}) with
\begin{equation}
    \mathcal{A}_{\rm UV} =  \mathcal{A}_{\rm UV,0} + \mathcal{A}_{\rm VL,1} + \mathcal{A}_{\rm VL,2} + \mathcal{A}_{\rm GB} + \mathcal{A}_{\rm R} \,,
\end{equation}
we obtain
\bea
 &&   [C_{H\ell}^{(3)}]_{\tau\tau}(\mu) =  -\frac{1}{16\pi^2}\frac{N_{\rm c}C_U}{2} \Big[ |W_{11}|^2|y_t|^2\left(1+\log\frac{\mu^2}{\mlq}\right) \nonumber\\
    && \qquad + c_Q^2 2\text{Re} (W_{11}^* W_{21} Y_+^* y_t) B_0(x_Q)  \no\\
    && \qquad  + c_Q^4 |W_{21}|^2 (|Y_+|^2 + |Y_-|^2) F(x_Q, x_Q^R) \Big] \,,
\eea
and 
\bea
 &&   [C_{H\ell}^{(3)}]_{\mu\mu}(\mu) =  -\frac{1}{16\pi^2}\frac{N_{\rm c}C_U}{2} s_L^2\Big[ |W_{12}|^2|y_t|^2\left(1+\log\frac{\mu^2}{\mlq}\right) \nonumber\\
    && \qquad  + c_Q^2 2\text{Re} (W_{12}^* W_{22} Y_+^* y_t) B_0(x_Q)  \no\\
    && \qquad  + c_Q^4 |W_{22}|^2 (|Y_+|^2 + |Y_-|^2) F(x_Q, x_Q^R) \Big]\,,
\eea
where $F(x_Q, x_Q^R)=B_1(x_Q) - B_2(x_Q) - \tan^2\beta B_2(x_Q^R)$.
Having introduced a single VL fermion, the first 
generation leptons do not couple to the new dynamics and $[C_{H\ell}^{(3)}]_{ee}=0$.

\subsection{Matching to $C_{\ell\ell}$}
\label{sect:Cll}

The one-loop (LQ-box) contributions to the SMEFT operator $[O_{\ell\ell}]_{\alpha\beta\gamma\delta}$ have been 
calculated in Ref.~\cite{Fuentes-Martin:2020hvc}. The coefficients relevant to our analysis 
are 
\be
    [C_{\ell\ell}]_{\tau\mu\mu\tau}  = 
    [C_{\ell\ell}]_{\mu\tau\tau\mu} = C_U \frac{g_U^2}{16\pi^2} s_L^2 B_{\ell\ell}^{1212}\,,
\ee
where the explicit expression for the functions $B_{\ell\ell}^{ijkl}$ can be found in Ref.~\cite{Fuentes-Martin:2020hvc}.
Also in this case, Wilson coefficients involving first generation fermions have vanishing contributions. 

So far, we assumed that the third-generation lepton doublet 
charged under $SU(4)$, namely $\ell_L^3$, can be identified with 
the $\ell_L^\tau$ doublet, defined by the $\tau$ mass-eigenstate. 
In general, a small misalignment is possible. 
If the RH current of the $U_1$ is close to its expectation in the minimal setup 
(i.e.~if $\beta_R^{b\tau}\approx 1$), bounds from $\tau \to \mu\gamma$ 
and $B_s \to \mu^+\mu^-$ allow a mixing of at most $O(10^{-2})$  between $\ell_L^3$ 
and the mass-eigenstate $\ell_L^\mu$~\cite{Fuentes-Martin:2019mun}. These bounds are less stringent 
if $|\beta_R^{b\tau}| \ll 1$: in this case the $\ell_L^3$--$\ell_L^\mu$ mixing,
that we parameterize  via the 
angle $s_\tau$ defined as in~\cite{Fuentes-Martin:2019mun}, 
could be as large as  $O(10^{-1})$.
A non-vanishing $s_\tau$, up to $O(10^{-1})$, has a negligible impact in
all the amplitudes evaluated so far. However, it leads to an additional non-vanishing 
contribution to $[C_{\ell\ell}]_{\tau\mu\mu\tau}$ via the 
tree-level $Z^\prime$-exchange amplitude (which involves only $\ell_L^3$).
Neglecting the subleading terms of $\mathcal{O}(g_{\rm SM}^2)$~\cite{Fuentes-Martin:2020hvc}, this contribution yields
\be
       [C_{\ell\ell}]_{\tau\mu\mu\tau} =   [C_{\ell\ell}]_{\mu\tau\tau\mu}
     = \frac{3g_U^2 v^2}{16 m_{Z'}^2} s_\tau^2~.
     \label{eq:Zp}
\ee

\section{Numerics}
We have now  all the ingredients to estimate the complete impact of 4321 dynamics on the  leptonic LFU ratios. 
Putting all the pieces together, the corrections to the leptonic decay widths defined in Eq.~(\ref{eq:dR}) 
assume the following form
\bea
&&   	\delta R_{\mu e} =  \frac{N_{\rm c}}{16\pi^2}C_U \Big[|\beta_L^{b\mu}|^2|y_t|^2\left(1+2L_U\right)  \no\\
&&\qquad 		+ 4 c_Q \text{Re} (\beta_L^{b\mu^*} \beta_L^{Q\mu\, } Y_+^* y_t) B_0(x_Q)  \no\\
&&\qquad  	+ 2 c_Q^2 |\beta_L^{Q\mu}|^2 (|Y_+|^2 + |Y_-|^2)F(x_Q, x_Q^R) \Big]\,, \\
&&  	\delta R_{\tau e} =  \frac{N_{\rm c}}{16\pi^2}C_U\Big[|\beta_L^{b\tau}|^2|y_t|^2\left(1+2L_U\right) \no \\
&&\qquad  	+  4 c_Q \text{Re} (\beta_L^{b\tau^*} \beta_L^{Q\tau\, } Y_+^* y_t) B_0(x_Q)  \no\\
&&\qquad 		+ 2 c_Q^2 |\beta_L^{Q\tau}|^2 (|Y_+|^2 + |Y_-|^2)F(x_Q, x_Q^R) \Big]\,,  \\
&&   \delta R_{\tau \mu}  =  \frac{N_{\rm c}}{16\pi^2}C_U\Big\{ (|\beta_L^{b\tau}|^2+|\beta_L^{b\mu}|^2)|y_t|^2\left(1+2L_U\right) \no\\
&&\qquad  	+ 4 c_Q  \text{Re}\left[\left(\beta_L^{b\tau^*} \beta_L^{Q\tau} + \beta_L^{b\mu^*} \beta_L^{Q\mu}\right) Y_+^* y_t\right] 
			B_0(x_Q)   \no\\
&&\qquad 		+ 2 c_Q^2 (|\beta_L^{Q\tau}|^2+|\beta_L^{Q\mu}|^2) (|Y_+|^2 + |Y_-|^2)F(x_Q, x_Q^R) \Big\}  \no \\
&&\qquad+ C_U\frac{g_U^2}{4\pi^2 } s_L^2 B_{\ell\ell}^{1212}  + \frac{3g_U^2 v^2}{4 x_{Z'}m_U^2}s^2_\tau\,,
\eea
where $L_U =\log (m_t^2/\mlq )$ and $x_{Z'} = m_{Z'}^2/m_U^2$.
 In the above expressions we have replaced the dependence from the 
$W_{ij}$ matrix elements via the  effective $U_1$ couplings  defined as in  Eq.~(\ref{eq:JUgen}):
\begin{align}\label{eq:betaDict}
\begin{aligned}
\beta_L^{b \tau} &= W_{11}\,,&
\beta_L^{b \mu}  &= -s_L W_{12}\,, \\
\beta_L^{Q \tau} &= c_Q W_{21}\,,&
\beta_L^{Q \mu}  &= -c_Q s_L W_{22}\,.
\end{aligned}
\end{align}

\begin{figure}[t]
\centering
\includegraphics[scale=0.55]{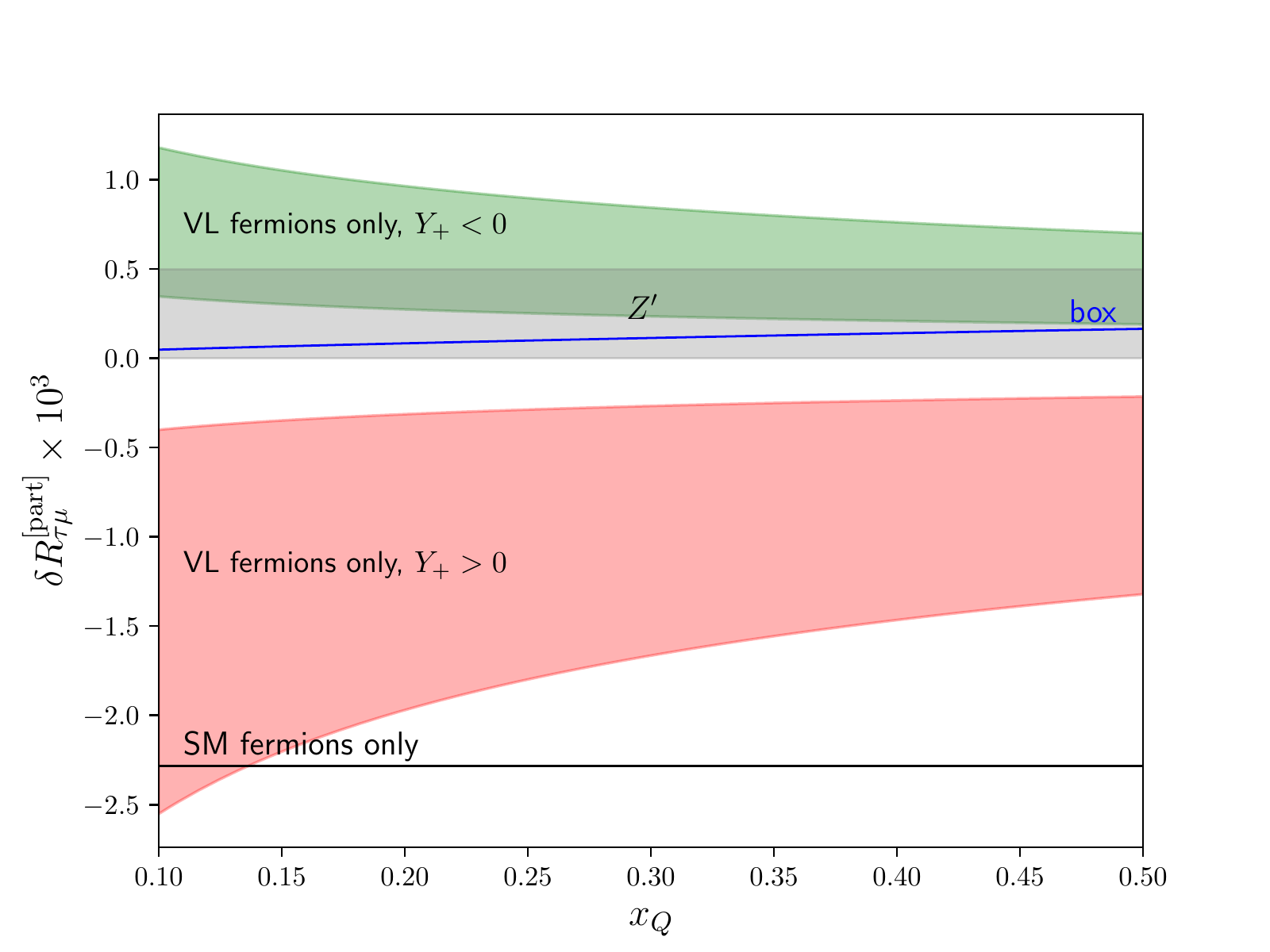} 
\caption{\label{fig:zoom} Separate contributions to $\delta R_{\tau \mu}$ in the 4321 model, 
using the benchmark values in Eqs.~(\ref{eq:bench1})--(\ref{eq:bench4}), as a function of $x_Q$: 
1) modified $W$-boson vertex 
from $U_1$+SM fermions (black line), 2) modified $W$-boson vertex from box with one and two VL fermions, 
for $Y_+ > 0$ (red) or $Y_+ < 0$ (green), 3)~$Z^\prime$-exchange amplitude  (grey band);
4) four-lepton box amplitude (blue line).}
\end{figure}

To estimate the size of the different contributions to the LFU ratios we choose the following benchmark 
values 
for the LQ couplings 
\begin{align}
 C_U  = \frac{g_U^2 v^2}{4 \mlq} &= 0.01\,, \quad  &  m_U = 4\ \rm TeV\,,\,  \no\\
 \beta_L^{b\tau} &= 1, \quad  &\beta_L^{b\mu} = - 0.2\,. \quad
  \label{eq:bench1}
\end{align}
These values are representative of the best-fit point obtained in Ref.~\cite{Cornella:2021sby}
when fitting present data, in the limit of small RH couplings ($|\beta_R^{b\tau}| \ll 1$),
and are consistent with the model expectation $W_{ij} =O(1)$ and $|s_L| \ll1$.
For the latter reason we further set 
\be
c_Q=1\,, \quad     \beta_L^{Q\tau} = 1, \quad  \  \beta^{Q\mu} = - 0.2\,.
  \label{eq:bench2}
\ee
Given the discussion in Section~\ref{sect:2B},
we neglect  $Y_-$ and vary $Y_+$ in the interval
\be
  0.2 < |Y_+|  < 1.0~,
   \label{eq:bench3}
\ee
corresponding to $|V_{cb}| < |s_Q| <  |V_{us}|$.
In principle, the signs of $c_Q$, $ \beta_L^{Q\tau}$, and  $ \beta^{Q\mu}$ could be varied;
however, only the relative sign of these couplings and $Y_+$ is relevant in $\delta R_{\alpha\beta}$. Therefore  
we effectively explore all relevant options varying the sign of $Y_+$.

Concerning the $Z^\prime$ contribution, at fixed $C_U$ 
the result in Eq.~(\ref{eq:Zp}) depends only on the combination 
$  s_\tau^2 / x_{Z'} $. For the sake of simplicity, we set 
$x_{Z'}=3/5$\footnote{Thos corresponds to a heavy mass for coloron, $m^2_G = (9/5) m^2_U$,
 which better evades direct constraints~\cite{Cornella:2021sby}.}
 and vary $s_\tau$ in the interval
\be
0 < s_\tau < 0.1~.
   \label{eq:bench4}
\ee
We finally consider the limit of heavy radial excitation, setting $x_Q^R = x_Q/(4\pi)^2$.

In Fig.~\ref{fig:zoom} we show all contributions separately in the case of  $\delta R_{\tau \mu}$, 
which is sensitive to all types of amplitudes. As expected, the contribution from
$U_1$+SM fermions, which includes the LL result, is dominant. However, the contribution from 
VL fermions represents a significant correction. On the other hand, the $Z^\prime$-exchange 
and the four-lepton box amplitudes are clearly subleading and safely negligible in most of the 
parameter space.\footnote{The $Z^\prime$-exchange amplitude is below $10\%$, in size, 
of the leading contribution from $U_1$+SM fermions for   $|s_\tau| \lsim 0.07$.} 

\begin{figure}[t]
\centering
\includegraphics[scale=0.5]{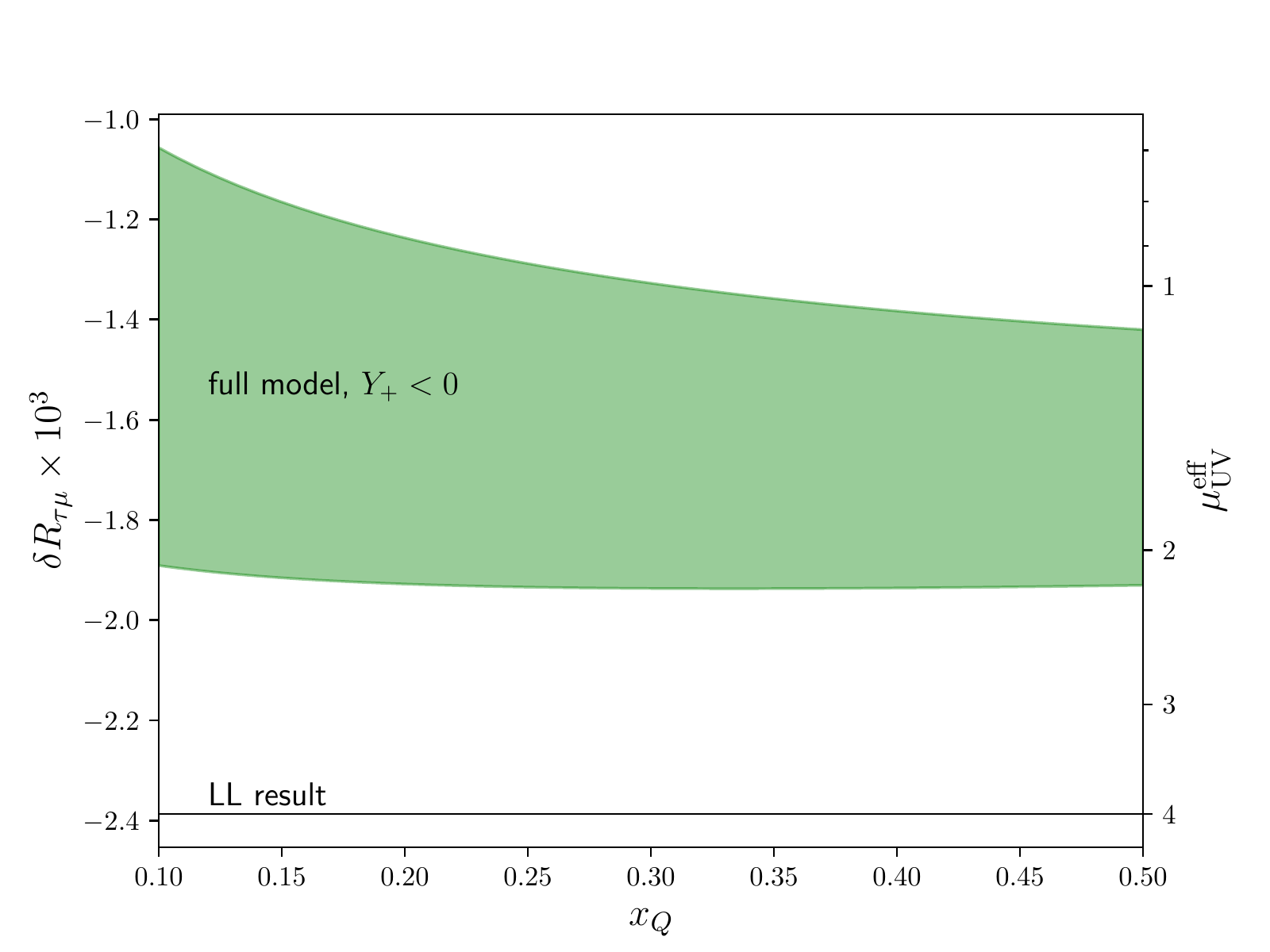} 
\caption{Comparison of the full calculation for $\delta R_{\tau \mu}$ (green band) in the case 
$Y_+ <0$, with the LL result obtained using $m_U=4$~TeV as UV matching scale (black line). 
The vertical scale on the right indicates the effective UV matching scale necessary for the 
LL result to produce a value of $\delta R_{\tau \mu}$  as indicated by the vertical scale on the left.}
\label{fig:mUeff}
\end{figure}

\begin{figure}[t]
\centering
	\includegraphics[scale=0.50]{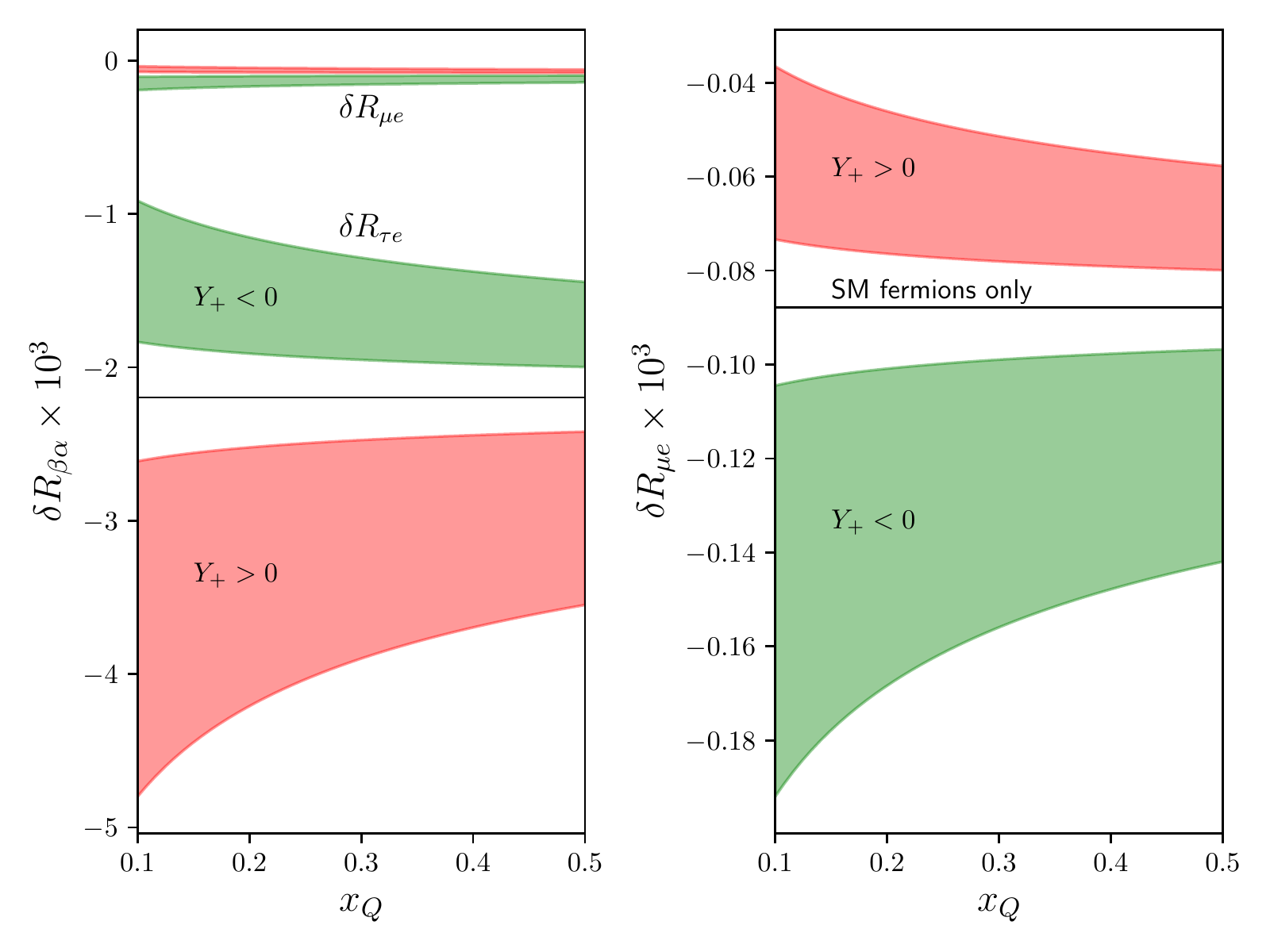} \\
\caption{Predictions for $\delta R_{\tau e}$ and $\delta R_{\mu e}$ in the 4321 model, using the benchmark values in 
Eqs.~(\ref{eq:bench1})--(\ref{eq:bench3}), as a function of $x_Q$.  The bands correspond to the full model results.
The right plot is a zoom on the upper part of the left plot in order to illustrate the different contributions to $\delta R_{\mu e}$.}
\label{fig:tauemue}
\end{figure}

In Fig.~\ref{fig:mUeff} we compare the LL result for $\delta R_{\tau \mu}$ with the full calculation in the case of $Y_+ <0$
(and $s_\tau =0$), 
where VL fermions decrease the effect induced by SM fermions only. As expected, in this case 
the effect is equivalent to that of decreasing the UV matching scale of the the LL result, 
from its natural value (namely $m_U$). The correction is sizable, corresponding to 
an effective decrease of the matching scale from 4~TeV to about 2~TeV or less. 
This effect is very relevant in decreasing the present tension with data when fitting 
the $B$ anomalies~\cite{Cornella:2021sby}.

In Fig.~\ref{fig:tauemue} we show the results for both $\delta R_{\tau e}$ and $\delta R_{\mu e}$.
As expected, the result for $\delta R_{\tau e}$ is almost identical to that of $\delta R_{\tau \mu}$, 
whereas the breaking of universality in $\delta R_{\mu e}$ is one order of magnitude smaller, 
reaching $\cO(10^{-4})$ at most. Note that in both cases, the unambiguous predictions following from 
$B$ anomalies is a reduction of the LFU ratios from one.
 
 \section{The effective $W$- and $Z$-boson couplings.}
 
\begin{figure}[t]
\centering
\includegraphics[scale=0.55]{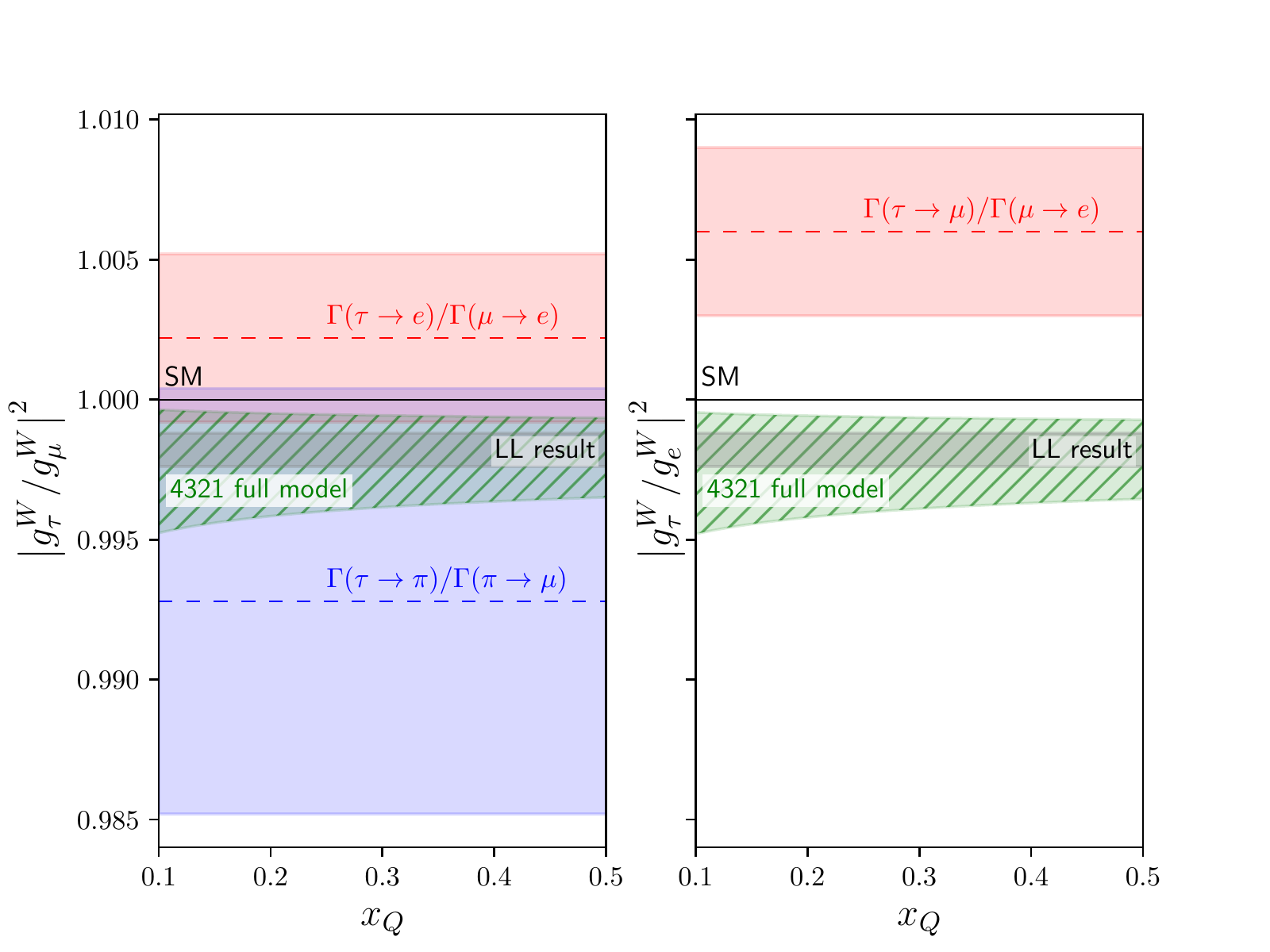} 
\caption{\label{fig:data}
Comparison between data and model predictions for the effective $W$-boson couplings to leptons ($g^W_\ell$).
The hatched green areas indicate the predictions of 4321 models (full model) in the region relevant to 
the $B$-physics anomalies. The corresponding LL results are highlighted in gray. 
The red bands are the experimental results obtained using purely leptonic decays~\cite{Pich:2013lsa}.
The blue band is the value of $|g^W_\tau / g^W_\mu|^2$
determined  in Ref.~\cite{Arroyo-Urena:2021nil}  from $\Gamma(\tau\to\pi\nu)/\Gamma(\pi\to\mu\bar\nu)$.}
\end{figure}
 
 The smallness of the $Z^\prime$-exchange 
and the four-lepton box amplitudes allow us to describe the breaking of universality in leptonic $\tau$ decays 
occurring in 4321 models 
as  modifications of the effective $W$-boson couplings to leptons ($g^W_\ell$).
Defining the latter as 
\be
{\cal{L}}^{(\ell,W)}_{\rm eff} =-{g_\ell^W \over\sqrt{2}}~ \overline{\nu}_\ell \gamma^{\mu}  P_L \ell ~W^+_{\mu} +{\rm h.c.},
\ee
the ratios introduced in Eq.~(\ref{eq:one}) can be expressed as
\be
 \left| \frac{g^{(\tau)}_e}{g^{(\mu)}_e}\right|^2  	\approx   \left| \frac{g_{\tau}^W}{g_\mu^W} \right|^2~.   
\ee
The smallness of NP effects for one-particle irreducible amplitudes 
  involving the first generation of quarks,
  implies the same effective $W$-boson couplings can also be extracted from   $\Gamma(\tau\to\pi\nu)$ and $\Gamma(\pi\to\mu\bar\nu)$.

 In Fig.~\ref{fig:data} we compare our results with the extraction of 
 $|g^W_{\tau}/{g^W_\mu}|$ using both leptonic  and pion decays:
 \bea
 \left| g_{\tau}^W/ g_{\mu}^W \right|^2_{\tau-\rm decays} &=& 1.0022\pm 0.0030~\text{\cite{Pich:2013lsa}}\,,  
 \label{Eq:gW1}
 \\
 \left| g_{\tau}^W/ g_{\mu}^W \right|^2_{\pi-\rm decays} &=&   0.9928\pm 0.0076 ~\text{\cite{Arroyo-Urena:2021nil}}\,.
  \label{Eq:gW2}
 \eea
 We also compare the model prediction for  $|g^W_{\tau}/{g^W_e}|$ with 
  \bea
 \left| g_{\tau}^W/ g_{e}^W \right|^2_{\tau-\rm decays} &=& 1.0060\pm 0.0030~\text{\cite{Pich:2013lsa}}\,. 
   \label{Eq:gW3}
 \eea
 In order to obtain robust estimates, we vary $C_U$
 in the interval $0.005 < C_U <0.01$ (with fixed $m_U=4$~TeV) and consider both $Y_+>0$ and $ Y_+<0$
 (with $s_\tau=0$). 
 As can be seen, in the $|g^W_{\tau}/{g^W_\mu}|$ case 
 present data are not precise enough to distinguish the SM from the 4321 model
 (in the region relevant to the $B$-physics anomalies).
 In the $|g^W_{\tau}/{g^W_e}|$ case, the inclusion of the contributions from VL fermions decreases the tension 
 with present data, which is reduced to about $2\sigma$ for $Y_+<0$. In both cases,
 a reduction of the present error by a factor 2-3 on the $\tau$ decay widths, 
 which might be accessible at Belle-II, could allow to perform very stringent test 
 of 4321 models in the motivated parameter-space region.

For completeness, we note that in this framework also the left-handed couplings 
of the $Z$ boson to charged leptons and neutrinos are modified.
The calculation proceeds very similarly to the one presented in Section~\ref{sect:matching}
for the $W$-boson couplings,  the only relevant difference being the presence 
of the singlet operator $O_{H\ell}^{(1)}$.
The  corresponding matching conditions reads
\bea
 &&   [C_{H\ell}^{(1)}]_{\tau\tau}(\mu) =  \frac{1}{16\pi^2}\frac{N_{\rm c}C_U}{2} \Big[ |W_{11}|^2|y_t|^2\left(1+\log\frac{\mu^2}{\mlq}\right) \nonumber\\
    && \qquad + c_Q^2 2\text{Re} (W_{11}^* W_{21} Y_+^* y_t) B_0(x_Q)  \no\\
    && \qquad  + c_Q^4 |W_{21}|^2 (|Y_+|^2 - |Y_-|^2) F(x_Q, x_Q^R) \Big] \,,
\eea
\bea
 &&   [C_{H\ell}^{(1)}]_{\mu\mu}(\mu) =  \frac{1}{16\pi^2}\frac{N_{\rm c}C_U}{2} s_L^2\Big[ |W_{12}|^2|y_t|^2\left(1+\log\frac{\mu^2}{\mlq}\right) \nonumber\\
    && \qquad  + c_Q^2 2\text{Re} (W_{12}^* W_{22} Y_+^* y_t) B_0(x_Q)  \no\\
    && \qquad  + c_Q^4 |W_{22}|^2 (|Y_+|^2 - |Y_-|^2) F(x_Q, x_Q^R) \Big]\,,
\eea
while $[C_{H\ell}^{(1)}]_{ee} \approx 0$.
Defining the effective left-handed $Z$-boson couplings as
\be
\mathcal{L}_{\text{eff}}^{(\ell,Z)} = - \frac{g_2}{c_W} \left[ g_{\ell_L}^Z (\bar \ell \gamma^\mu P_L \ell)  + g_{\nu_\ell}^Z (\bar \nu_\ell \gamma^\mu  P_L \nu_\ell ) \right] Z_\mu\,, 
\ee
where $g_2$ is the $SU(2)_L$ gauge coupling, $c_W$ denotes the cosine of the weak angle, 
and  $g^{Z,\text{SM}}_{\nu_\ell} = -  g^{Z,\text{SM}}_{\ell_L} = 
1/2$, the modified couplings ($g^Z_i = g^{Z,\text{SM}}_i + \delta g^Z_i$) are
\bea
\delta g^Z_{\nu_\ell}(\mu) &=& - \frac{v^2}{2} \left\{ [C_{H\ell}^{(1)}]_{\ell\ell}(\mu) -  [C_{H\ell}^{(3)}]_{\ell\ell}(\mu)\right\}\,, \label{eq:gZ1} \\
\delta g^Z_{\ell_L}(\mu) &=& - \frac{v^2}{2} \left\{ [C_{H\ell}^{(1)}]_{\ell\ell}(\mu) +  [C_{H\ell}^{(3)}]_{\ell\ell}(\mu)\right\} \,.   \label{eq:gZ2}
\eea
Since the leading contributions controlled by $y_t$ and $Y_+$ are equal and opposite in 
$C_{H\ell}^{(1)}$ and $C_{H\ell}^{(3)}$, Eqs.~(\ref{eq:gZ1})--(\ref{eq:gZ2}) 
imply a sizable modification of $g^Z_{\nu_\tau}$  and negligible corrections to all the $g^Z_{\ell_L}$. 
Neglecting the subleading contribution 
proportional to $|Y_-|^2$ we get 
\be
\left.
\delta g^Z_{\ell_L} \right|_{Y_-=0} =  0\,,    \quad 
\left. \frac{ \delta g^Z_{\nu_\ell}}{ g^{Z,{\rm SM}}_{\nu_\ell} } \right|_{Y_-=0}
=
\left. 
\frac{ \delta g^W_{\ell}}{ g^{W,{\rm SM}}_{\ell} }\right|_{Y_-=0}.
\ee
According to this result, the most significant constraint on the model 
from $Z$-pole observables arises by the  invisible decay width of the $Z$-boson,
or the effective number of LH neutrinos ($N_\nu^{\rm eff}$) determined by this observable~\cite{ALEPH:2005ab}.
Assuming that only $\delta g^Z_{\nu_\tau}$ receives a sizable correction (as expected in our model),
we find
\be
\left| \frac{  g^Z_{\nu_\tau}}{ g^{Z,\rm SM}_{\nu_\tau} }  \right|^2_{ N_\nu^{\rm eff} } = 
N_\nu^{\rm eff} - 2 =  0.9840 \pm  0.0082~,
\ee
which is slightly less stringent than the constraints from the effective $W$ couplings in 
Eqs.~(\ref{Eq:gW1})--(\ref{Eq:gW3}).

\section{Conclusion}

The recent $B$-physics anomalies have strengthened the importance of precise tests of LFU
in all accessible processes involving charged leptons. In this paper we have presented the first complete analysis of LFU 
violations in 
 leptonic $\tau$ decays, within the motivated class of 4321 models addressing the $B$-physics anomalies~\cite{DiLuzio:2017vat,Bordone:2017bld,Greljo:2018tuh,DiLuzio:2018zxy,Fuentes-Martin:2020bnh,Fuentes-Martin:2020hvc,Fuentes-Martin:2020pww}.
As originally pointed out in Ref.~\cite{Feruglio:2016gvd,Feruglio:2017rjo} via a general EFT approach,
the $b\to c \tau\nu$ anomaly implies a decrease of the effective $W$-boson coupling to $\tau$ leptons in the 
few per-mil range. While confirming this general conclusion, we have shown that the inclusion of vector-like fermions, 
which is motivated by $B$-physics data in this context, can lead to sizable modifications of the EFT results.
In particular, the inclusion of vector-like fermions can partially decrease the present tension 
of 4321 models with data on leptonic decays.
Most importantly, the results presented in this work  could  lead to very stringent tests for 
this class of models, in the region favored by $B$-physics data, with the help of 
future precision measurements of  leptonic $\tau$ decay widths.

\subsection*{Acknowledgments}

We thank Claudia Cornella and Javier Fuentes-Mart\'{\i}n  for useful discussions. 
This project has received funding from the European Research Council (ERC) under the European Union's Horizon 2020 research and innovation programme under grant agreement 833280 (FLAY), and by the Swiss National Science Foundation (SNF) under contract 200021-175940. 

\bibliographystyle{jhep}
\bibliography{lfu_tau}

\end{document}